\documentclass{aa}

\usepackage{graphicx}
\usepackage{txfonts}

\DeclareUnicodeCharacter{00A0}{~}

\newcommand{\gammaSI}{\,J\,m$^{-2}$\,s$^{-0.5}$\,K$^{-1}$}
\newcommand{\radd}{\text{rad}\,\text{d}^{-2}}
\newcommand{\dd}{\mathrm{d}}

\begin{document}

\title{YORP and Yarkovsky effects in asteroids (1685)~Toro, (2100)~Ra-Shalom, (3103)~Eger, and (161989)~Cacus}

\titlerunning{YORP and Yarkovsky effects in asteroids (1685)~Toro, (2100)~Ra-Shalom, (3103)~Eger, and (161989)~Cacus}

\author{J.~\v{D}urech           \inst{1} \and
        D.~Vokrouhlick\'y       \inst{1} \and
        P.~Pravec               \inst{2} \and
        J.~Hanu\v{s}            \inst{1} \and
        D.~Farnocchia           \inst{3} \and
        Yu.~N.~Krugly           \inst{4} \and
        V.~R.~Ayvazian          \inst{5} \and
        P.~Fatka                \inst{1,2} \and
        V.~G.~Chiorny           \inst{4} \and
        N.~Gaftonyuk            \inst{4} \and
        A.~Gal\'ad              \inst{2} \and
        R.~Groom                \inst{6} \and
        K.~Hornoch              \inst{2} \and
        R.~Y.~Inasaridze        \inst{5} \and
        H.~Ku\v{c}\'akov\'a     \inst{1,2} \and
        P.~Ku\v{s}nir\'ak       \inst{2} \and
        M.~Lehk\'y              \inst{1} \and
        O.~I.~Kvaratskhelia     \inst{5} \and
        G.~Masi                 \inst{7} \and
        I.~E.~Molotov           \inst{8} \and
        J.~Oey                  \inst{9} \and
        J.~T.~Pollock           \inst{10} \and
        V.~G.~Shevchenko        \inst{4} \and   
        J.~Vra\v{s}til          \inst{1} \and
        B.~D.~Warner            \inst{11}
       }

\institute{
     Institute of Astronomy, Faculty of Mathematics and Physics, Charles University,
     V Hole\v{s}ovi\v{c}k\'ach 2, 18000, Prague, Czech Republic\\
     \email{durech@sirrah.troja.mff.cuni.cz}
\and 
     Astronomical Institute, Czech Academy of Sciences, Fri\v{c}ova 298, Ond\v{r}ejov, Czech Republic
\and 
    Jet Propulsion Laboratory, California Institute of Technology, Pasadena, CA 91109, USA
\and 
    Institute of Astronomy of Kharkiv National University, Sumska Str.~35, 61022 Kharkiv, Ukraine
\and 
    Kharadze Abastumani Astrophysical Observatory, Ilia State University, K. Cholokoshvili Av. 3/5, Tbilisi 0162, Georgia
\and 
    Darling Range Observatory, Perth, WA, Australia
\and 
    Physics Department, University of Rome ``Tor Vergata'', Via della Ricerca Scientifica 1, 00133 Rome, Italy 
\and 
    Keldysh Institute of Applied Mathematics, RAS, Miusskaya 4, Moscow 125047, Russia
\and 
    Blue Mountains Observatory, 94 Rawson Pde. Leura, NSW 2780, Australia
\and 
    Physics and Astronomy Department, Appalachian State University, 525 Rivers St, Boone, NC 28608, USA
\and 
    Center for Solar System Studies -- Palmer Divide Station, 446 Sycamore Ave., Eaton, CO 80615, USA
}

\date{Received ???; accepted ???}
 
\abstract
{The rotation states of  small asteroids are affected by a net torque arising from an anisotropic sunlight reflection and thermal radiation from the asteroids' surfaces. On long timescales, this so-called YORP effect can change asteroid spin directions and their rotation periods.}
{We analyzed lightcurves of four selected near-Earth asteroids with the aim of detecting secular changes in their rotation rates that are caused by YORP or at least of putting upper limits on such changes.}
{We use the lightcurve inversion method to model the observed lightcurves and include the change in the rotation rate $\dd \omega / \dd t$ as a free parameter of optimization. To enlarge the time line of observations and to increase the sensitivity of the method, we collected more than 70 new lightcurves. For asteroids Toro and Cacus, we used thermal infrared data from the WISE spacecraft and estimated their size and thermal inertia by means of a thermophysical model. We also used the currently available optical and radar astrometry of Toro, Ra-Shalom, and Cacus to infer the Yarkovsky effect.
} 
{We detected a YORP acceleration of $\dd\omega / \dd t = (1.9 \pm 0.3) \times 10^{-8}\,\radd$ for asteroid Cacus. The current astrometric data set  is not sufficient to provide detection of the Yarkovsky effect in this case. 
For Toro, we have a tentative ($2\sigma$) detection of YORP from a significant improvement of the lightcurve fit for a nonzero value of $\dd\omega / \dd t = 3.0 \times 10^{-9}\,\radd$. We note an excellent agreement between the observed secular change of the semimajor axis $\dd a / \dd t$ and the theoretical expectation for densities in the 2--2.5\,g\,cm$^{-3}$ range.
For asteroid Eger, we confirmed the previously published YORP detection with more data and updated the YORP value to $(1.1 \pm 0.5) \times 10^{-8}\,\radd$. 
We also updated the shape model of asteroid Ra-Shalom and put an upper limit for the change of the rotation rate to $|\dd\omega / \dd t| \lesssim 1.5 \times 10^{-8}\,\radd$. Ra-Shalom has a greater than $3\sigma$ Yarkovsky detection with a theoretical value consistent with observations assuming its size and/or density is slightly larger than the nominally expected values. 
Using the convex shape models and spin parameters reconstructed from lightcurves, we computed theoretical YORP values and compared them with those measured. They agree with each other within the expected uncertainties of the model.}
{}


\maketitle

\section{Introduction}

The rotation state of small ($\lesssim 30$\,km) asteroids can be affected on long timescales by a net torque that is caused by directly scattered sunlight and thermal radiation from the surfaces of the asteroids. This so-called YORP effect can change the directions of the rotation axis and the rotation rates \citep{Bot.ea:06, Vok.ea:15} and has direct consequences for the distribution of asteroid rotation periods \citep{Pra.ea:08} and obliquities \citep{Han.ea:13b, Pra.ea:12}. The YORP effect is also believed to be the driving mechanism for creating asteroid binaries and pairs by rotation fission \citep{Pra.ea:10, Marg.ea:15}. 

In a similar vein, orbits of small asteroids are affected by the Yarkovsky effect,  a net reaction force from the thermal radiation of a rotating body with nonzero thermal inertia; it has crucial consequences for the evolution of the main asteroid belt, for the supply of the near-Earth asteroid (NEA) population, and for impact hazard assessment \citep[see the review by][]{Vok.ea:15}. The Yarkovsky drift has been detected for more than one hundred near-Earth asteroids \cite[e.g.,][]{Far.ea:13,Gre.ea:17}.

While the evolution of the spin axis cannot be detected from current photometric data, a change in the rotation rate can be detected because time-resolved photometry is very sensitive to even a small secular change in the rotation period. So far, YORP-driven acceleration of the rotation period has been directly detected in five asteroids \citep[see the review by][]{Vok.ea:15} and there are indirect detections of YORP-driven evolution of spins of members of asteroid families \citep{Vok.ea:03, Car.ea:16, Pao.Kne:16}. Additional direct detections are needed if we want to compare real values of YORP with those predicted by theoretical models \citep[][for example]{Roz.Gre:13, Gol.ea:14, Low.ea:14, Sev.ea:15}. The YORP-driven evolution of asteroid rotation plays a crucial role in the dynamical evolution of the whole asteroid population and only new measurements of the YORP effect together with theoretical models will enable us to create a self-consistent model of this process.

To enlarge the sample of asteroids with a YORP detection, we analyzed  archival lightcurves and new data of four NEAs that, according to the estimated YORP magnitude, should have a detectable deviation from the constant-period rotation. We also reevaluate observation constraints for the Yarkovsky effect and put them into context with the size and thermal inertia we derived for our targets.

\section{YORP detection through lightcurve inversion}
\label{sec:models}

To look for possible secular changes in the rotation period, we used the lightcurve inversion method of \cite{Kaa.ea:01}. \cite{Kaa.ea:03} slightly modified the method so that it included one more free parameter in the optimization: the change of the rotation rate $\upsilon = \dd\omega / \dd t$. We applied this modified lightcurve inversion to archived photometric lightcurves (references given below) and our new observations (Tables 1--4). For each asteroid, we reconstructed its convex shape model and tested whether a nonzero $\upsilon$ value provides a significantly better fit to the data than a model with constant period ($\upsilon = 0$). We estimated the uncertainties of the derived parameters with the same approach as \cite{Dur.ea:12b} from the $\chi^2$ distribution with a given degrees of freedom. If not stated otherwise, the reported uncertainties are $1\sigma$. 

\subsection{(3103) Eger}
\label{sec:Eger}

\begin{figure}[t]
\begin{center}
\includegraphics[width=\columnwidth]{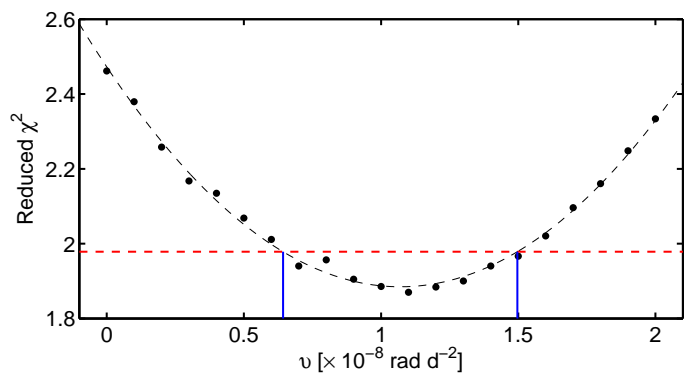}
\caption{\label{fig_Eger_YORP_uncerainty} Dependence of the goodness of the fit measured by the reduced $\chi^2$ on the YORP parameter $\upsilon$ for asteroid Eger. The best fit is for $\upsilon = 1.1 \times 10^{-8}\,\radd$. The dashed curve is a quadratic fit of the data points. The dashed red line indicates a $6\%$ increase in the $\chi^2$, which defines our $3\sigma$ uncertainty interval given the number of degrees of freedom.}
\end{center}
\end{figure}

This is one of five asteroids in which YORP has been detected. \cite{Dur.ea:12b} determined the YORP acceleration to $\upsilon = (1.4 \pm 0.6) \times 10^{-8}\,\radd$ ($3\sigma$ error), and included  a ``warning'' that data from upcoming apparitions would be needed to confirm this detection. By adding more observations from 2014 and 2016 \citep{War.ea:17} and our two lightcurves from 2017 (see Table~\ref{table_Eger}), we confirmed previous results and derived an updated value $\upsilon = (1.1 \pm 0.5) \times 10^{-8}\,\radd$ ($3\sigma$ error) with slightly better precision. For a realistic estimate of the uncertainty interval, we used the same approach as \cite{Vok.ea:11, Vok.ea:17} by assuming that the $3\sigma$ uncertainty interval is defined by all solutions with $\chi^2 < (1 + 3\sqrt{2\nu})\, \chi^2_\text{min} $, where $\chi^2_\text{min}$ was the $\chi^2$ of the best model and $\nu$ was the number of degrees of freedom ($\nu \sim 5500$ in case of Eger). The dependence of $\chi^2$ on the YORP parameter $\upsilon$ is shown in Fig.~\ref{fig_Eger_YORP_uncerainty} for $\upsilon$ between 0 and $2.0 \times 10^{-8}\,\radd$. The $3\sigma$ uncertainty corresponds to an  increase in $\chi^2$ of about 6\%. The new convex shape model is very similar to that published by \cite{Dur.ea:12b}. However, we recall that to significantly decrease the uncertainty of the $\upsilon$ value, it is necessary to extend the data arc with observations from the next apparitions.

\begin{table}[tb]
\caption{\label{table_Eger}     
 Aspect data for new observations of (3103) Eger.}
\centering
\begin{tabular}{cccrcrc}
\hline \hline
 Date & $r$  & $\Delta$ & \multicolumn{1}{c}{$\alpha$} & $\lambda$ & \multicolumn{1}{c}{$\beta$}  & Obs. \\
      & [AU] & [AU]     & [deg]    & [deg]     & [deg]                        &  \\ 
\hline  
 2017 02 05.8  & 1.487    & 0.523  & 13.6     & 152.8     & $13.6$ & Ond \\
 2017 02 16.1  & 1.534    & 0.564  & 11.4     & 146.3     & $17.8$ & Ond \\
\hline
\end{tabular}
\tablefoot{
 The table lists Eger's distance from the Sun $r$ and from the Earth
 $\Delta$, the solar phase angle $\alpha$, the geocentric ecliptic coordinates of
 the asteroid $(\lambda, \beta)$, and the observatory (Ond -- Ond\v{r}ejov, 65 cm).}
\end{table}

\subsection{(1685) Toro}
\label{sec:Toro}

\begin{figure}[t]
\begin{center}
\includegraphics[width=\columnwidth]{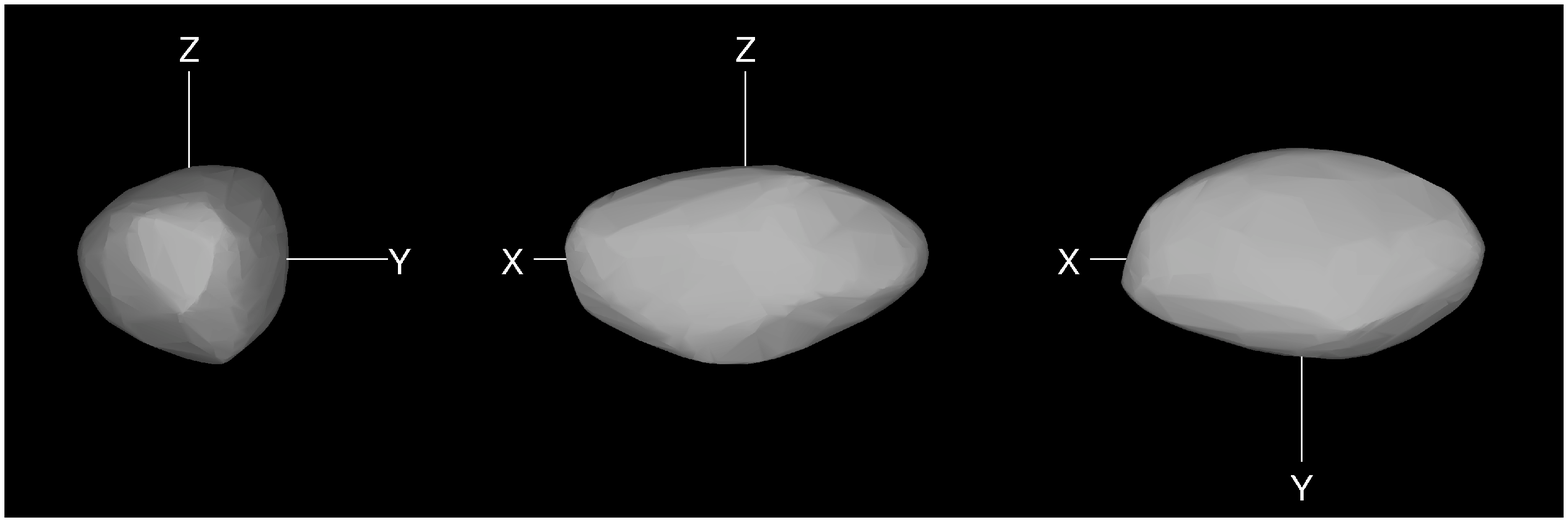}
\caption{\label{fig_Toro_shape}
  Shape model of (1685)~Toro shown from equatorial level (left and center, $90\degr$ apart) and pole-on (right).}
\end{center}
\end{figure}

\begin{figure*}[t]
\begin{center}
\includegraphics{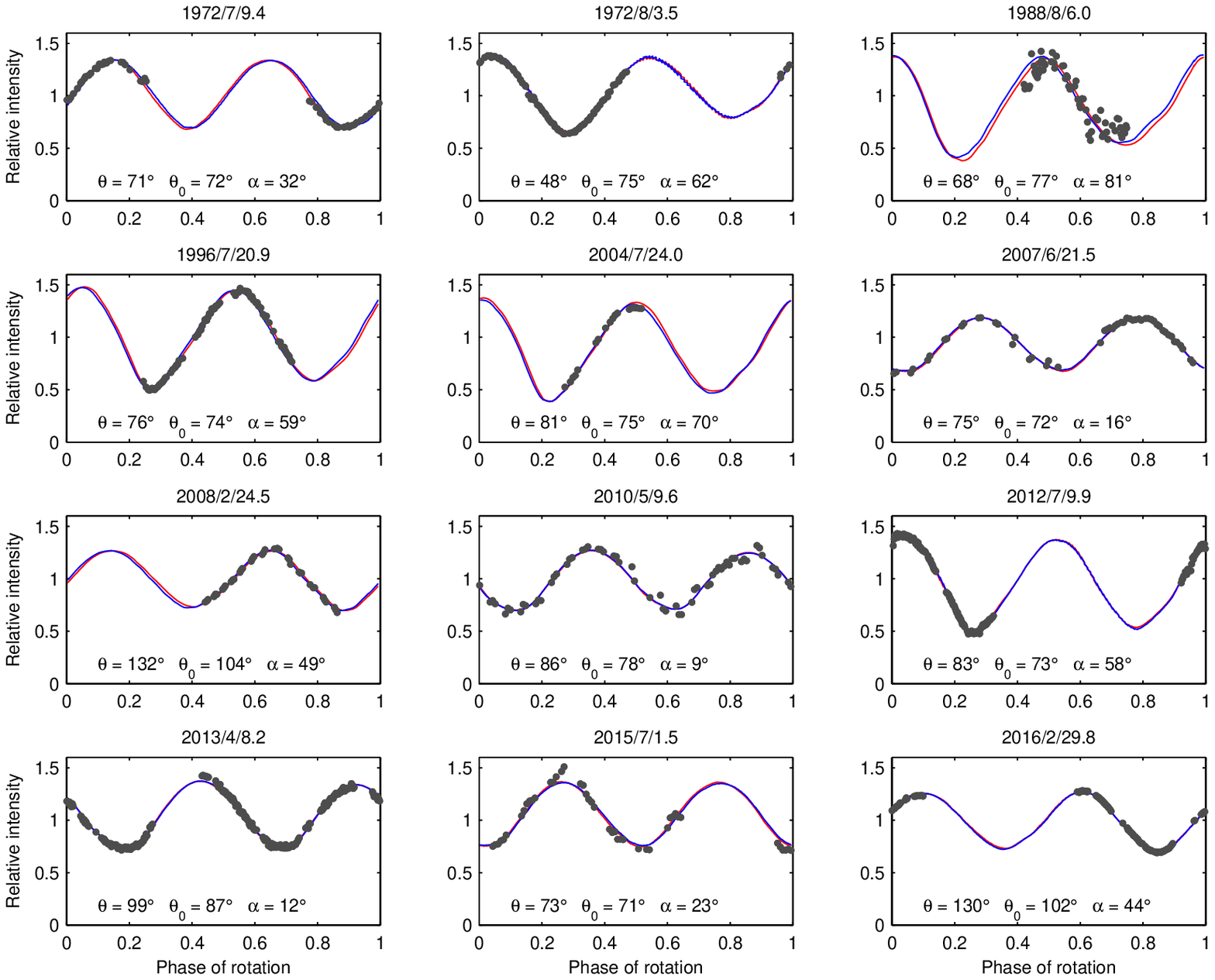}
\caption{\label{fig_Toro_lc}
 Example lightcurves of (1685)~Toro shown with the synthetic lightcurves produced by the best-fit constant-period model (blue) and with YORP (red). The geometry is described by the aspect angle $\theta$, the solar aspect angle $\theta_0$, and the solar phase angle $\alpha$.}
\end{center}
\end{figure*}

For Toro we used archived lightcurves from 1972 \citep{Dun.ea:73}, 1998 \citep{Hof.Gey:90}, 2007 \citep{Hig:08}, 2008 \citep{Hig.ea:08}, 2010 \citep{Hig:11,Oey:11}, and 2013 \citep{War:13}, and we also observed new lightcurves (see Table~\ref{table_Toro}). From this data set, we reconstructed the shape model (Fig.~\ref{fig_Toro_shape}), the rotation period $P = (10.19782 \pm 0.00003$)\,hr for 8.5 July 1972 (the date of the first photometric observation),
the pole direction $(\lambda, \beta) = (71 \pm 10^\circ, -69 \pm 5^\circ)$ (corresponding to obliquity $\epsilon = 161 \pm 6^\circ$), and $\upsilon = 3.0 \times 10^{-9}\,\radd$. For this value of $\upsilon$, the $\chi^2$ drops by 11\% with respect to $\chi^2$ for $\upsilon = 0$. The formal phase shift over the interval of 44 years corresponding to this value of $\upsilon$ is only $22^\circ$. The difference between the constant period and YORP model is most pronounced for lightcurves from 1996. They are also crucial for YORP detection. When  the four lightcurves observed in 1996 are removed, the difference between the YORP model with the best-fit value $\upsilon = 2.3 \times 10^{-9}\,\radd$ and a constant period model with  $\upsilon = 0$ is only 4\% in $\chi^2$. To confirm this tentative YORP detection, additional data from future apparitions are needed. The fit to the selected lightcurves is shown in Fig.~\ref{fig_Toro_lc}.

\begin{figure}[t]
\begin{center}
\includegraphics[width=\columnwidth]{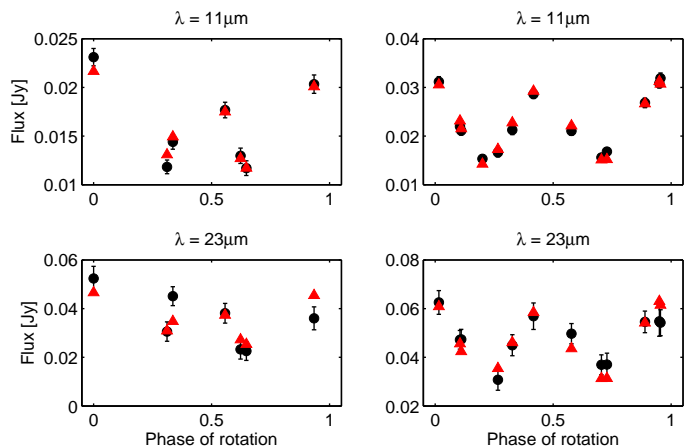}
\caption{\label{fig_Toro_IR}
  Comparison between the model (red triangles) and Toro thermal infrared data observed by WISE on 10 February 2010 (left), and 15 July 2010 (right).}
\end{center}
\end{figure}

To better characterize this asteroid, we also used observations of the Wide-field Infrared Survey Explorer (WISE) satellite \citep{Wri.ea:10, Mai.ea:11b}. WISE observed Toro in two epochs (10 February and 15 July) in 2010, the data from the W3 (11\,$\mu$m) and W4 (23\,$\mu$m) filters are available through the IRSA/IPAC archive. We checked the data against the quality and reliability criteria described in \cite{Ali.ea:16}. Using our shape model and spin parameters, we applied the thermophysical model of \cite{Lag:96a, Lag:97, Lag:98} to derive the thermophysical properties. The best fit with the reduced $\chi^2 = 1.4$ is for thermal inertia $260^{+140}_{-110}$\,\gammaSI, high roughness, and albedo $0.13 \pm 0.03$, assuming the values $H = 13.9$\,mag and $G = 0.11$ from the database of asteroid absolute magnitudes and slopes \citep{Mui.ea:10,Osz.ea:11}. The size of the asteroid is $3.5^{+0.3}_{-0.4}$\,km, which is in good agreement with the mean effective diameter of $\sim 3.3$\,km derived by \cite{Ost.ea:83} from radar observations. The fit to the data is shown in Fig.~\ref{fig_Toro_IR}.

\begin{table}[tb]
\caption{\label{table_Toro}     
 Aspect data for new or unpublished observations of (1685) Toro.}
\centering
\begin{tabular}{cccrcrc}
\hline \hline
 Date & $r$  & $\Delta$ & \multicolumn{1}{c}{$\alpha$} & $\lambda$ & \multicolumn{1}{c}{$\beta$}  & Obs. \\
      & [AU] & [AU]     & [deg]    & [deg]     & [deg]                        &  \\ 
\hline  
1996 07 16.0  & 1.155    & 0.278  & 54.0     & 358.8     & $20.7$ & Sim \\
1996 07 16.9  & 1.149    & 0.273  & 54.9     &   0.4     & $21.5$ & Sim \\
1996 07 17.0  & 1.148    & 0.272  & 55.0     &   0.6     & $21.5$ & Sim \\
1996 07 20.9  & 1.120    & 0.251  & 59.6     &   8.5     & $25.0$ & Sim \\
2004 07 24.0  & 1.077    & 0.273  & 69.8     &  25.1     & $24.8$ & Kh \\
2004 09 13.1  & 0.787    & 0.525  & 98.2     & 121.1     & $13.6$ & Kh \\
2012 07 07.9  & 1.174    & 0.379  & 56.7     &   0.3     & $14.4$ & Kh \\
2012 07 08.9  & 1.167    & 0.372  & 57.4     &   1.9     & $15.0$ & Kh \\
2012 07 09.0  & 1.167    & 0.372  & 57.5     &   1.9     & $15.0$ & Kh \\
2012 07 09.9  & 1.160    & 0.366  & 58.3     &   3.5     & $15.5$ & Kh \\
2012 07 24.0  & 1.060    & 0.306  & 73.4     &  31.5     & $22.6$ & Si \\
2012 07 30.0  & 1.018    & 0.302  & 80.9     &  45.9     & $24.4$ & Kh \\
2012 07 30.9  & 1.011    & 0.302  & 82.1     &  48.2     & $24.6$ & Kh \\
2012 08 04.0  & 0.983    & 0.308  & 86.9     &  57.9     & $24.8$ & Ab \\
2013 04 07.2  & 1.809    & 0.839  & 11.6     & 190.6     & $-20.2$ &    Pro \\
2013 04 08.2  & 1.812    & 0.843  & 11.8     & 190.1     & $-20.2$ &    Pro \\
2015 07 01.5  & 1.650    & 0.736  & 23.3     & 239.4     & $-4.3$ &      BMO \\
2015 07 03.5  & 1.640    & 0.741  & 24.8     & 238.6     & $-4.0$ &        BMO \\
2015 07 04.6  & 1.635    & 0.743  & 25.6     & 238.3     & $-3.8$ &        BMO \\
2015 07 05.4  & 1.631    & 0.746  & 26.2     & 238.0     & $-3.7$ &        BMO \\
2015 07 06.5  & 1.626    & 0.748  & 26.9     & 237.7     & $-3.5$ &      BMO \\
2016 01 22.5  & 0.975    & 0.157  & 88.6     &  23.5     & $-19.3$ &        DRO \\
2016 02 04.6  & 1.066    & 0.200  & 61.2     &  65.9     & $-27.4$ &        DRO \\
2016 02 05.6  & 1.073    & 0.206  & 59.8     &  68.4     & $-27.5$ &        DRO \\
2016 02 08.6  & 1.095    & 0.226  & 56.0     &  75.2     & $-27.4$ &        DRO \\
2016 02 09.6  & 1.102    & 0.233  & 54.9     &  77.2     & $-27.3$ &        DRO \\
2016 02 10.6  & 1.109    & 0.240  & 53.9     &  79.2     & $-27.2$ &        DRO \\
2016 02 11.6  & 1.116    & 0.247  & 52.9     &  81.0     & $-27.0$ &        DRO \\
2016 02 12.6  & 1.123    & 0.255  & 52.0     &  82.8     & $-26.8$ &        DRO \\
2016 02 13.6  & 1.130    & 0.263  & 51.2     &  84.4     & $-26.7$ &        DRO \\
2016 02 14.5  & 1.137    & 0.270  & 50.5     &  85.9     & $-26.5$ &        DRO \\
2016 02 17.6  & 1.159    & 0.295  & 48.4     &  90.3     & $-25.8$ &        DRO \\
2016 02 22.2  & 1.191    & 0.335  & 46.1     &  95.9     & $-24.8$ &     PDO \\
2016 02 23.2  & 1.198    & 0.344  & 45.7     &  96.9     & $-24.6$ &     PDO \\
2016 02 24.2  & 1.205    & 0.353  & 45.4     &  98.0     & $-24.3$ &     PDO \\
2016 02 24.8  & 1.210    & 0.359  & 45.1     &  98.6     & $-24.2$ & Ab \\
2016 02 25.2  & 1.213    & 0.363  & 45.0     &  99.0     & $-24.1$ &     PDO \\
2016 02 26.3  & 1.220    & 0.373  & 44.6     & 100.0     & $-23.9$ &     PDO \\
2016 02 26.8  & 1.224    & 0.378  & 44.5     & 100.5     & $-23.7$ & Ab \\
2016 02 27.7  & 1.230    & 0.387  & 44.2     & 101.4     & $-23.5$ & Ab \\
2016 02 28.2  & 1.233    & 0.391  & 44.1     & 101.8     & $-23.4$ &     PDO \\
2016 02 29.8  & 1.244    & 0.406  & 43.7     & 103.2     & $-23.1$ & Ab \\
2016 03 16.9  & 1.352    & 0.575  & 41.2     & 114.6     & $-19.7$ &      BE \\
2016 04 05.7  & 1.475    & 0.809  & 40.0     & 125.7     & $-16.4$ & Si \\
2016 04 06.7  & 1.481    & 0.821  & 39.9     & 126.2     & $-16.3$ & Ab \\
2016 04 09.8  & 1.498    & 0.860  & 39.8     & 127.8     & $-15.8$ & Ab \\
2016 05 02.8  & 1.620    & 1.162  & 38.2     & 139.1     & $-13.1$ & Ab \\
2016 05 04.7  & 1.630    & 1.188  & 38.0     & 140.0     & $-12.9$ & Ab \\
2016 05 08.8  & 1.649    & 1.243  & 37.7     & 142.0     & $-12.5$ & Ab \\
\hline
\end{tabular}
\tablefoot{
 The table lists Toro's distance from the Sun $r$ and from the Earth
 $\Delta$, the solar phase angle $\alpha$, the geocentric ecliptic coordinates of
 the asteroid $(\lambda, \beta)$, and the observatory or source (DK -- Danish telescope, La Silla, $1.54$ m; BMO -- Blue Mountains Observatory, 35 cm, BE -- Blue Eye, Ond\v{r}ejov, 60 cm; Si -- Simeiz Observatory, Crimea, 1 m; Ab -- Abastumani Observatory, Georgia, 70 cm; Kh -- Astronomical Observatory, Kharkiv, 70 cm; Pro -- Prompt Observatory, Cerro Tololo, 41 cm; PDO -- Palmer Divide Observatory, 35 cm; DRO -- Darling Range Observatory, 30 cm).}
\end{table}

\subsection{(161989) Cacus}
\label{sec:Cacus}

The first lightcurves of Cacus come from 1978 observations of \cite{Sch.ea:79} and \cite{Deg.ea:78}. During 2003, the asteroid was observed from Ond\v{r}ejov observatory; \cite{Koe.ea:14} observed one lightcurve in 2009; and we observed this asteroid in 2014--16 at La Silla. The whole set covers 20 years and five apparitions (see Table~\ref{table_Cacus}). On 17 February 2015 we measured the color index in the Johnson-Cousins photometric system $(V-R) = (0.486 \pm 0.015)$\,mag.  From observations taken on 8 and 15 December 2015, we derived the mean absolute magnitude $H = (17.51 \pm 0.19)$\,mag assuming the phase slope parameter $G = 0.24 \pm 0.11$, which is the mean $G$ value for S- and Q-type asteroids \citep{Pra.ea:12b}.

\begin{figure*}[t]
\begin{center}
\includegraphics{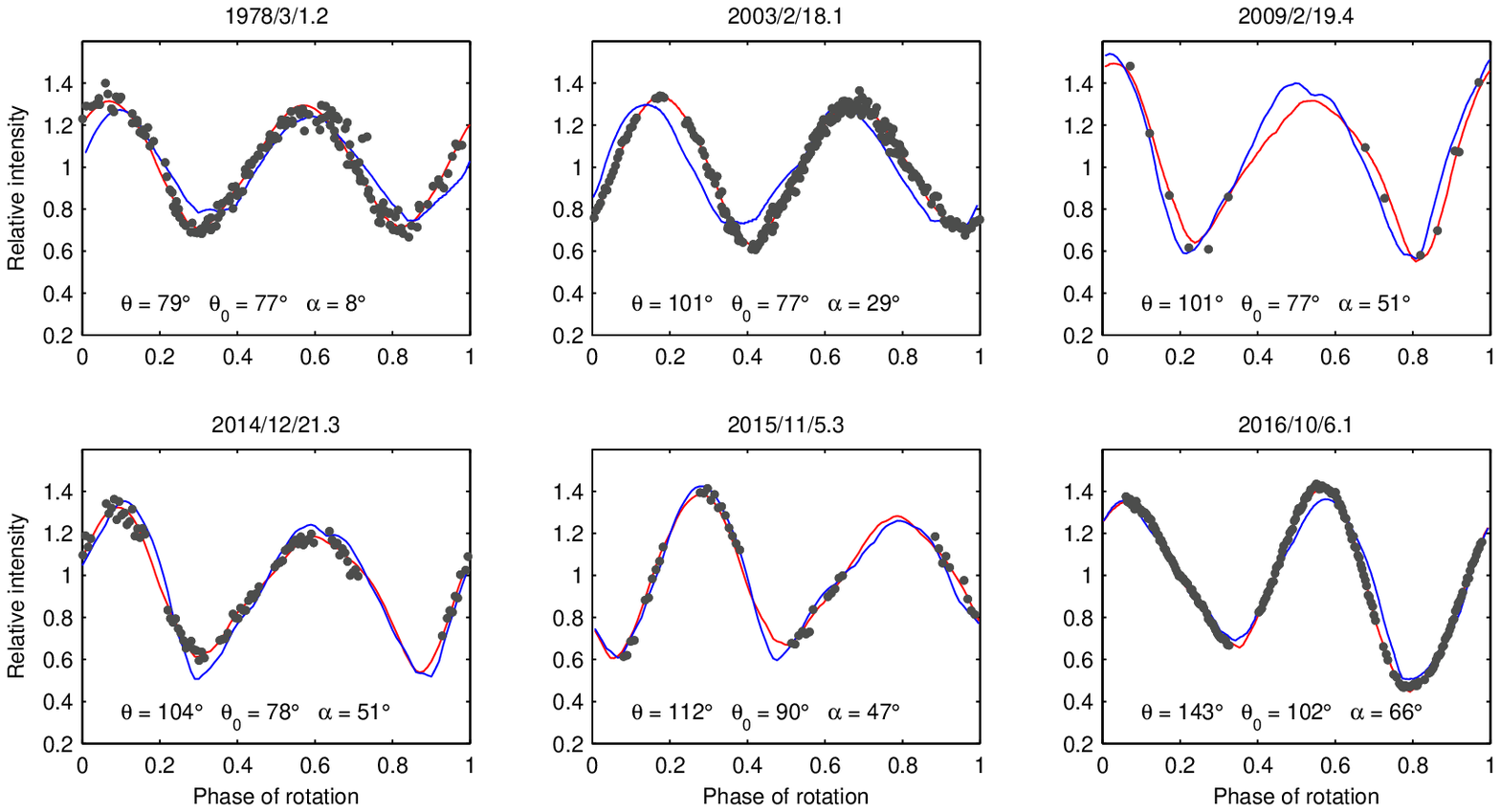}
\caption{\label{fig_Cacus_lc}
 Example lightcurves of (161989)~Cacus shown with the synthetic lightcurves produced by the best-fit constant-period model (blue) and with YORP (red). The geometry is described by the aspect angle $\theta$, the solar aspect angle $\theta_0$, and the solar phase angle $\alpha$.}
\end{center}
\end{figure*}

\begin{figure}[t]
\begin{center}
\includegraphics[width=\columnwidth]{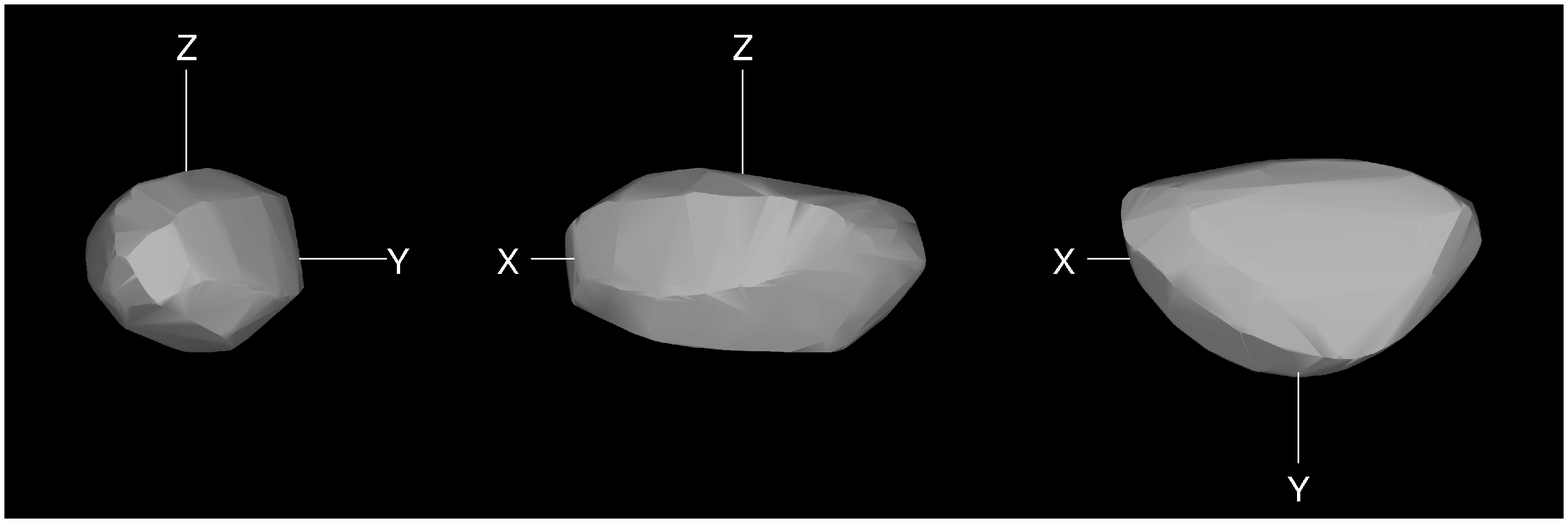}
\caption{\label{fig_Cacus_shape}
  Shape model of (161989)~Cacus shown from equatorial level (left and center, $90\degr$ apart) and pole-on (right).}
\end{center}
\end{figure}

We applied the lightcurve inversion to the photometric data set and derived a unique shape model. The fit to lightcurves with $\upsilon = 0$ was not satisfactory, but if we allowed the rotation rate to change, we got a significantly better fit (see Fig.~\ref{fig_Cacus_lc}). Our final model has the pole direction $\lambda = (254 \pm 5)^\circ$, $\beta = (-62 \pm 2)^\circ$ (corresponding obliquity is $\epsilon = 178 \pm 3^\circ$) and rotation period $P = (3.755067 \pm 0.000002)$\,hr for 28.5 February 1978. The best value for the change in the rotation rate is $\upsilon = (1.9 \pm 0.3) \times 10^{-8}\,\radd$. The shape model is shown in Fig.~\ref{fig_Cacus_shape}.

\begin{figure}[t]
\begin{center}
\includegraphics[width=\columnwidth]{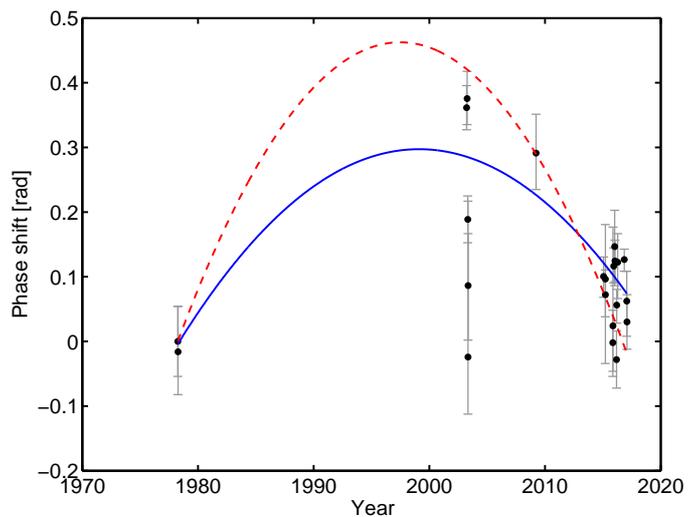}
\caption{\label{fig_Cacus_phase_shift} 
  Phase shift between the best constant-period model of Cacus and the observed lightcurves. Each point in the plot represents a single lightcurve, the error bars represent the uncertainty of the phase shift given the number of points and the level of noise in each lightcurve. The blue curve is the best least-squares quadratic fit to the data taking into account the error bars. The red dashed curve is the quadratic phase shift corresponding to the best YORP model derived with lightcurve inversion.}
\end{center}
\end{figure}

In Fig.~\ref{fig_Cacus_phase_shift}, we show the phase shift between the best constant-period model ($P = 3.755054$\,hr) and the real data. For each observed lightcurve, we created a corresponding smooth synthetic lightcurve produced by the best-fit constant-period model and then computed the phase shift of synthetic data that produced the best match between the two lightcurves. These values are shown in the plot together with the error bars estimated from the number of points and the level of noise in each lightcurve. If YORP changes the rotation rate, this $\mathrm{O} - \mathrm{C}$ difference should be a quadratic function of time. The trend is not very clear mainly because of three lightcurves from 2003, but they have large error bars  and a small number of points, so their contribution is less significant. The formal YORP coefficient obtained by fitting a second-order polynomial to the phase-shift points in Fig.~\ref{fig_Cacus_phase_shift} is $1.0\times 10^{-8}\,\radd$.

The detection of the YORP acceleration is critically dependent on the first two lightcurves from 1978. If we exclude them from the data set, the observations span only 13 years and the difference between the constant period and YORP model is not statistically significant: both models provide essentially the same fit to the data and the phase offset between the models for $\upsilon = 1.9 \times 10^{-8}\,\radd$ is only $\sim 3^\circ$. However, because the two lightcurves were obtained by independent observers and instruments \citep{Deg.ea:78, Sch.ea:79}, it is not likely that they were  both  shifted in time the same way to mimic the YORP effect.

\begin{figure}[t]
\begin{center}
\includegraphics[width=\columnwidth]{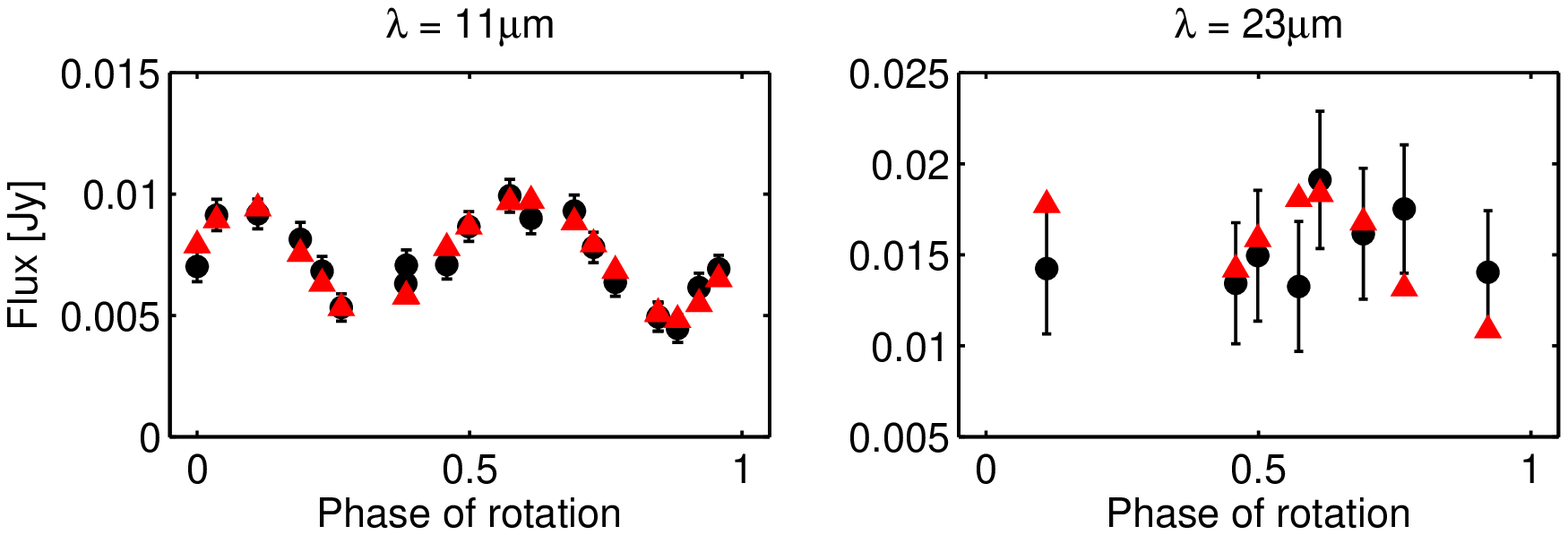}
\caption{\label{fig_Cacus_IR}
  Comparison between the model (red triangles) and Cacus thermal data observed by WISE on 14 February 2010.}
\end{center}
\end{figure}

Similarly to Toro, we also used observations of the WISE spacecraft, which observed Cacus in 2010. Using our shape model and spin parameters, we applied the thermophysical model and derived the thermophysical properties. Because the thermal data in W4 filter have large uncertainties, they only provide loose constraints to the thermal inertia and the size of the asteroid. The best fit with the reduced $\chi^2 = 0.7$ is for a thermal inertia around 500--800\,\gammaSI, but all of the values in the range 250--2000\,\gammaSI with low to medium surface roughness provide a very good fit to the data (Fig.~\ref{fig_Cacus_IR}). The modeled size of Cacus is $(1.0 \pm 0.2)\,$km, which gives the  albedo $p_V = 0.18 \pm 0.08$. 

\begin{table}[tb]
\caption{\label{table_Cacus}    
 Aspect data for all available observations of (161989) Cacus.}
\centering
\begin{tabular}{cccrcrc}
\hline \hline
 Date & $r$  & $\Delta$ & \multicolumn{1}{c}{$\alpha$} & $\lambda$ & \multicolumn{1}{c}{$\beta$}  & Obs. \\
      & [AU] & [AU]     & [deg]    & [deg]     & [deg]                        &  \\ 
\hline  
1978 03 01.2  & 1.131    & 0.142  &  8.4     & 153.5     & $-6.6$ & S79 \\
1978 03 08.3  & 1.107    & 0.128  & 26.0     & 142.1     & $14.8$ & D78 \\
2003 02 18.1  & 1.186    & 0.232  & 28.5     & 143.5     & $-34.5$ & DK \\  
2003 03 05.8  & 1.134    & 0.181  & 35.2     & 124.1     & $-7.6$ & Mod \\ 
2003 03 25.9  & 1.063    & 0.230  & 67.3     & 106.9     & $30.0$ & Ond \\
2003 04 01.8  & 1.039    & 0.262  & 74.1     & 103.3     & $38.0$ & Ond \\
2003 04 04.9  & 1.028    & 0.276  & 76.5     & 101.9     & $40.8$ & Ond \\
2009 02 19.4  & 1.121    & 0.238  & 51.0     & 212.5     & $ 0.6$ & K14 \\
2014 12 21.3  & 1.253    & 0.904  & 51.2     & 186.4     & $-18.0$ & DK \\
2015 02 17.3  & 1.067    & 0.466  & 67.6     & 241.7     & $13.5$ & DK \\
2015 02 17.4  & 1.067    & 0.466  & 67.6     & 241.7     & $13.5$ & DK \\
2015 02 20.3  & 1.056    & 0.458  & 68.9     & 245.8     & $16.3$ & DK \\
2015 10 09.3  & 1.300    & 0.808  & 50.2     & 103.6     & $-40.8$ & DK \\
2015 10 13.4  & 1.308    & 0.799  & 49.6     & 105.7     & $-42.5$ & DK \\
2015 11 05.3  & 1.343    & 0.745  & 46.6     & 115.3     & $-51.9$ & DK \\
2015 12 08.3  & 1.363    & 0.656  & 42.5     & 113.2     & $-63.5$ & DK \\
2015 12 15.3  & 1.363    & 0.638  & 41.8     & 108.4     & $-64.8$ & DK \\
2016 02 04.2  & 1.309    & 0.577  & 44.6     &  75.6     & $-45.7$ & DK \\
2016 02 12.0  & 1.294    & 0.585  & 46.5     &  76.2     & $-39.8$ & DK \\
2016 03 10.1  & 1.226    & 0.644  & 53.9     &  84.5     & $-19.7$ & DK \\
2016 10 06.1  & 1.089    & 0.344  & 66.0     & 289.2     & $-23.1$ & DK \\
2016 12 22.1  & 1.310    & 0.952  & 48.4     & 354.6     & $-34.8$ & DK \\
2016 12 31.1  & 1.326    & 1.015  & 47.4     &   1.3     & $-33.8$ & DK \\
\hline
\end{tabular}
\tablefoot{
 The table lists Cacus's distance from the Sun $r$ and from the Earth
 $\Delta$, the solar phase angle $\alpha$, the geocentric ecliptic coordinates of
 the asteroid $(\lambda, \beta)$, and the observatory or source (S79 -- \cite{Sch.ea:79}; D78 -- \cite{Deg.ea:78}; K14 -- \cite{Koe.ea:14}, Ond -- Ond\v{r}ejov observatory, 65 cm; DK -- Danish telescope, La Silla, $1.54$ m; Mod -- Modra observatory, 60 cm).}
\end{table}

\subsection{(2100) Ra-Shalom}
\label{sec:RaShalom}

The modeling of Ra-Shalom aiming at YORP detection was done by \cite{Dur.ea:12b} who used the data set from 1978 to 2009. We extended the set with lightcurves from  two additional apparitions in 2013 and 2016 (see Table~\ref{table_RaShalom}). On 8 October 2016 we measured the color index in the Johnson-Cousins photometric system $(V-R) = (0.398 \pm 0.010)$\,mag. There is still no detectable signal of a YORP torque;  the YORP model provides virtually the same $\chi^2$ as the constant period model. However, with the enlarged time line, we were able to reduce the $3\sigma$ uncertainty interval of YORP to $-1.0 \times 10^{-8} < \upsilon < 1.5 \times 10^{-8}\,\radd$, which is about 2--3 times tighter  than in \cite{Dur.ea:12b}. Because the data set is dominated by the more recent lightcurves, the discrepancy between the data and the model are most pronounced for the 1978 and 1981 lightcurves. We also updated the spin pole orientation and decreased its uncertainty \citep[compared to the rather large uncertainty of pole direction in][]{Dur.ea:12b}: $\lambda = (292 \pm 15)^\circ$, $\beta = (-65 \pm 10)^\circ$, $P = (19.8200 \pm 0.0003)$\,hr, $\epsilon = 166 \pm 12^\circ$.

\begin{table}[tb]
\caption{\label{table_RaShalom} 
 Aspect data for new observations of (2100) Ra-Shalom.}
\centering
\begin{tabular}{cccrcrc}
\hline \hline
 Date & $r$  & $\Delta$ & \multicolumn{1}{c}{$\alpha$} & $\lambda$ & \multicolumn{1}{c}{$\beta$}  & Obs. \\
      & [AU] & [AU]     & [deg]    & [deg]     & [deg]                        &  \\ 
\hline  
2013 09 07.1  & 1.153    & 0.402  & 59.2     &  63.4     & $-10.7$ & Ond \\
2013 09 08.1  & 1.150    & 0.395  & 59.3     &  64.2     & $-11.3$ & Ond \\
2013 09 10.0  & 1.144    & 0.380  & 59.7     &  65.7     & $-12.7$ & Ond \\
2013 09 27.1  & 1.078    & 0.274  & 66.8     &  84.1     & $-28.8$ & Ond \\
2013 09 28.1  & 1.073    & 0.269  & 67.6     &  85.6     & $-30.0$ & Ond \\
2016 08 10.4  & 1.191    & 0.486  & 57.2     &  38.9     & $ 6.4$ &     PDO \\
2016 08 11.4  & 1.192    & 0.479  & 57.0     &  39.2     & $ 6.1$ &     PDO \\
2016 08 12.4  & 1.192    & 0.472  & 56.7     &  39.5     & $ 5.8$ &     PDO \\
2016 08 13.4  & 1.193    & 0.465  & 56.4     &  39.7     & $ 5.5$ &     PDO \\
2016 08 14.4  & 1.194    & 0.458  & 56.2     &  40.0     & $ 5.2$ &     PDO \\
2016 08 15.4  & 1.194    & 0.451  & 55.9     &  40.3     & $ 4.8$ &     PDO \\
2016 08 16.4  & 1.195    & 0.444  & 55.6     &  40.6     & $ 4.5$ &     PDO \\
2016 08 17.4  & 1.195    & 0.437  & 55.3     &  40.8     & $ 4.2$ &     PDO \\
2016 08 19.4  & 1.195    & 0.423  & 54.8     &  41.3     & $ 3.4$ &     PDO \\
2016 08 20.4  & 1.195    & 0.415  & 54.4     &  41.5     & $ 3.0$ &     PDO \\
2016 08 26.0  & 1.194    & 0.375  & 52.6     &  42.7     & $ 0.6$ & Ond \\
2016 08 28.0  & 1.193    & 0.361  & 51.8     &  43.0     & $-0.4$ & Ond \\
2016 08 30.1  & 1.191    & 0.346  & 51.0     &  43.4     & $-1.5$ & Ond \\
2016 08 31.0  & 1.190    & 0.339  & 50.6     &  43.5     & $-2.0$ & Ond \\
2016 09 03.0  & 1.187    & 0.318  & 49.3     &  43.8     & $-3.9$ & Ond \\
2016 09 10.1  & 1.176    & 0.270  & 45.9     &  44.1     & $-9.4$ & Ond \\
2016 09 11.7  & 1.172    & 0.259  & 45.1     &  44.0     & $-11.0$ &        BMO \\
2016 09 16.7  & 1.161    & 0.228  & 42.5     &  43.4     & $-16.5$ &        BMO \\
2016 09 19.8  & 1.153    & 0.210  & 41.2     &  42.6     & $-20.6$ &        BMO \\
2016 09 22.7  & 1.145    & 0.195  & 40.2     &  41.5     & $-25.2$ &        BMO \\
2016 09 23.7  & 1.141    & 0.190  & 40.0     &  41.0     & $-26.9$ &        BMO \\
2016 09 25.7  & 1.135    & 0.181  & 39.8     &  39.8     & $-30.6$ &        BMO \\
2016 09 26.7  & 1.132    & 0.177  & 39.9     &  39.1     & $-32.5$ &        BMO \\
2016 09 27.7  & 1.128    & 0.173  & 40.2     &  38.3     & $-34.5$ &        BMO \\
2016 10 08.3  & 1.086    & 0.150  & 51.5     &  20.4     & $-58.1$ & DK \\
2016 10 09.1  & 1.082    & 0.150  & 52.9     &  18.0     & $-59.7$ & DK \\
2016 10 10.2  & 1.077    & 0.150  & 55.0     &  14.0     & $-62.0$ & DK \\
2016 10 13.6  & 1.061    & 0.152  & 61.7     & 358.0     & $-67.7$ &        BMO \\
2016 10 14.6  & 1.055    & 0.154  & 63.9     & 351.6     & $-69.0$ &        BMO \\
2016 10 15.5  & 1.051    & 0.155  & 65.7     & 346.0     & $-69.9$ &        BMO \\
2016 10 17.6  & 1.039    & 0.159  & 70.0     & 331.6     & $-71.0$ & BMO \\
2016 10 25.7  & 0.993    & 0.183  & 85.1     & 287.8     & $-66.8$ & BMO \\
2016 10 26.6  & 0.987    & 0.186  & 86.8     & 284.7     & $-65.8$ & BMO \\
\hline
\end{tabular}
\tablefoot{
 The table lists Ra-Shalom's distance from the Sun $r$ and from the Earth
 $\Delta$, the solar phase angle $\alpha$, the geocentric ecliptic coordinates of
 the asteroid $(\lambda, \beta)$, and the observatory or source (Ond -- Ond\v{r}ejov observatory, 65 cm; DK -- Danish telescope, La Silla, $1.54$ m; BMO -- Blue Mountains Observatory, 35 cm; PDO -- Palmer Divide Observatory, 35 cm).}
\end{table}

\section{Comparison with the theoretical model}
Here we estimate how the detected change in rotation rate
for (161989)~Cacus, its tentative value for (1685)~Toro, and
limits set in the (2100)~Ra-Shalom case agree with the theoretical
expectations from the YORP effect. The analysis of (3103)~Eger was already presented by \cite{Dur.ea:12b}. We also reevaluate the observational constraints of the Yarkovsky effect for these asteroids and compare them with our model. This is not a straightforward
task. First, it requires
a thermophysical model of the analyzed asteroid, which
depends on a number of poorly known parameters. Second,
it has been recognized that the YORP strength is
sensitive to small-scale irregularities of the
asteroid shape, which are far beyond the resolution of our
coarse convex models \citep[e.g.,][and
references therein]{Vok.ea:15}. In this situation we decided to
adopt the simplest possible model and leave the door open
for further improvements in the future. As a consequence, the real uncertainties of our theoretical predictions are 
larger than the formal ones corresponding to the uncertainty of the input parameters.

In particular, we use the one-dimensional heat diffusion
model of \cite{Cap.Vok:04} and \cite{Cap.Vok:05}.
This approach is able to treat the
self-shadowing of surface facets. However, this capability is not implemented in
our computation where we only use a coarse convex model from
lightcurve inversion techniques. We treat each facet
independently and the time-dependent heat diffusion
propagates to the depth below the surface. We assumed that the core is isothermal and treat the surface boundary condition in its nonlinear form \citep[e.g.,][]{Cap.Vok:04, Cap.Vok:05}.
The formulation requires setting the
values of the surface thermal conductivity $K$,
density $\rho_{\rm s}$, and heat capacity $C$. These
quantities are also traditionally combined into the
surface thermal inertia $\Gamma=\sqrt{K \rho_{\rm s} C}$.
If the surface is known down to small scales (centimeters
to decimeters), there are more complications to
affect the YORP strength: (i) shadowing and mutual
thermal irradiation of the surface facets \citep[e.g.,][]{Roz.Gre:12, Roz.Gre:13} 
and (ii) thermal communication
of the surface facets \citep[e.g.,][]{Gol.Kru:12, Sev.ea:16}. While these effects have 
limited influence on the accuracy of the global thermal
acceleration (the Yarkovsky effect), they can significantly
change the global thermal torque (the YORP effect). Overall, 
the self-irradiation tends to decrease the magnitude of the
YORP effect \citep[e.g.,][]{Roz.Gre:13}, while the thermal
communication of the surface facets introduces a systematic
trend that accelerates the rotation instead of decreasing it, as caused by the YORP effect
\citep[e.g.,][]{Gol.Kru:12}. So the
typical mismatch arising from the simplified approach using
a smooth, convex shape model is  overprediction of
the YORP strength, and  apparent symmetry in
acceleration and deceleration of the rotation rate
\citep[e.g.,][]{Cap.Vok:04}. Because the
thermal effects take place only in the thin surface layer
(typically $\leq 1$\,m), the isothermal core bulk density
$\rho_{\rm b}$ is the next parameter that needs to be
specified. Finally, our model requires an equivalent
size $D$ of the asteroid (i.e., diameter of a sphere with
the same volume as the asteroid), the orientation of the spin
axis, and the rotation period. The last two are taken from our
solution in Sect.~\ref{sec:models}, and the size is estimated from infrared observations or radar data.

Once the thermal model converges to the solution of the
surface temperature distribution at any time during
the revolution about the Sun, we can use the converged solution to compute
both thermal force and torque. By numerically averaging
over one revolution about the Sun, we derive estimates
 of the Yarkovsky and YORP effects. The former
is represented with a single parameter, namely the secular
change in semimajor axis $(\dd a / \dd t)_{\rm mod}$, and the latter
with the secular change in the rotation rate
$\upsilon_{\rm mod}= (\dd\omega / \dd t)_{\rm mod}$. The YORP effect
in obliquity is too small to be directly observed and is not
reported here. The Yarkovsky $(\dd a / \dd t)_{\rm mod}$ value 
depends on the surface thermal inertia value $\Gamma$,
while the YORP $\upsilon_{\rm mod}$ value does not \citep[e.g.,][]{Cap.Vok:04}. 
We note that
our model does not include the YORP component due to the
directly reflected sunlight in optical band. This
would require additional unconstrained parameters to be
set \citep[e.g.,][]{Bre.Vok:11}. Given the rather
small value of the surface albedo, we consider this approximation
at the level of neglected effects of surface self-irradiation
or lateral thermal communication of the surface facets
mentioned above.

Finally, we mention straightforward scaling rules for
the size $D$ and bulk density $\rho_{\rm b}$ \citep[e.g.,][]{Vok.ea:15}:
(i) $(\dd a / \dd t)_{\rm mod}\propto 1/
(\rho_{\rm b} D)$ and (ii) $\upsilon_{\rm mod}\propto 1/
(\rho_{\rm b} D^2)$. In what follows we use their nominal
values, but our results can be easily recalibrated.

\subsection{(1685) Toro}
We used our nominal values $D=3.5$\,km, $(\lambda,\beta)=
(71^\circ,-69^\circ),$ and $P=10.19782$\,hr from Sect.~\ref{sec:Toro}, and
the bulk density of $2.5$\,g\,cm$^{-3}$, appropriate for the
S-type spectral classification of this body. A sufficiently
large range of the the thermal inertia values $\Gamma$ was
also scanned to see dependence of the Yarkovsky effect
on this parameter.

For the YORP effect, we obtained a nominal value
$\upsilon_{\rm mod} \simeq 10.9\times 10^{-9}\,\radd$. This
is somewhat larger, factor $\simeq 3.6$, than the
observationally hinted value $\upsilon \simeq 3.0\times 10^{-9}\,\radd$,
though the latter is quite uncertain in this case.
This is
the expected level of mismatch due to approximations mentioned
above.
Nevertheless, there is again a consistency in the possible
acceleration of the rotation rate.
\begin{figure}[pt]
\begin{center}
 \includegraphics[width=\columnwidth]{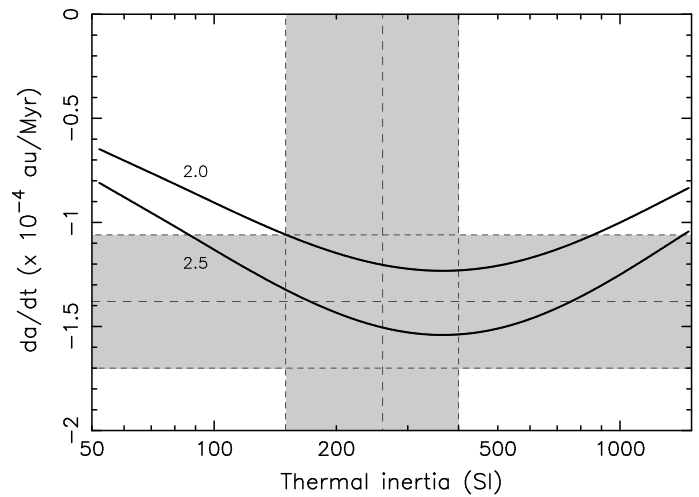}
\end{center}
\caption{Computed secular change of the orbital semimajor axis
 $(\dd a / \dd t)_{\rm mod}$ due to the Yarkovsky effect vs surface thermal
 inertia $\Gamma$ for (1685)~Toro. Nominal values of the
 rotation state and size are used. The gray vertical zone
 indicates the plausible range of $\Gamma$. The horizontal gray area is the Yarkovsky
 $\dd a / \dd t$ secular value from the orbit determination ($1\sigma$
 interval). The Yarkovsky effect has been clearly detected
 in spite Toro's large size due to radar astrometric observations
 in four apparitions. The theoretical curves for bulk densities
 $2$ and $2.5$\,g\,cm$^{-3}$ closely match  the overlap of the
 observed $\dd a / \dd t$ and $\Gamma$ values.}
\label{toro_dadt}
\end{figure}

Toro has been fortuitously observed by radar on several of
its close approaches to the Earth and also has a very long arc
of $68$ years over which the optical astrometry has been
collected. Therefore, in spite of this asteroid's large size, the Yarkovsky
effect has been detected fairly well. Our optical data revision, and 
radar astrometry from January 2016 added to the data set, yield 
$\dd a / \dd t = -(1.38\pm 0.32)
\times 10^{-4}$\,au\,Myr$^{-1}$. This is in a very good agreement
with the predicted $(\dd a / \dd t)_{\rm mod}$ value shown in
Fig.~\ref{toro_dadt}. The bulk density between $2$ and $2.5$\,g\,cm$^{-3}$ is the expected value for the S-type
spectral classification of this body.

\subsection{(161989) Cacus}
We used our nominal values $D=1$\,km, $(\lambda,\beta)=
(254^\circ,-62^\circ)$, and $P=3.755067$\,hr from Sect.~\ref{sec:Cacus}.
The uncertainties on the pole position and rotation period
have a negligible effect on our results. We also set a
bulk density $\rho_{\rm b}=2.5$\,g\,cm$^{-3}$, appropriate
for the Q-type classification of this body \citep{Tho.ea:14, Sch.ea:15}, and sampled the surface thermal
inertia range from Sect.~\ref{sec:Cacus}. 
With this set of parameters, we obtained $\upsilon_{\rm mod} \simeq
4.5\times 10^{-8}\,\radd$. While consistently predicting
an acceleration of the rotation rate, our theoretical value is about
a factor $\simeq 2.4$ higher than the observed value. 
\begin{figure}[pt]
\begin{center}
 \includegraphics[width=\columnwidth]{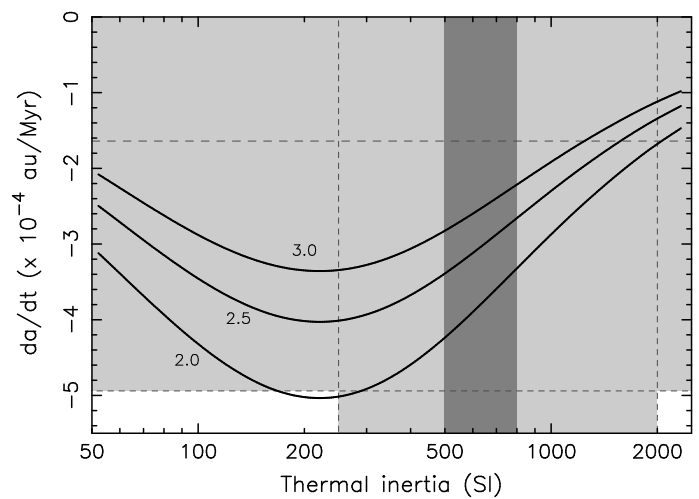}
\end{center}
\caption{Same as in Fig.~\ref{toro_dadt}, but for asteroid
 (161989)~Cacus. The solid curves show the theoretical dependence for three
 values of the bulk density $\rho_{\rm b}=2$\,g\,cm$^{-3}$, $2.5$\,g\,cm$^{-3}$, and $3$\,g\,cm$^{-3}$ (see the labels). The gray vertical zone
 indicates plausible range of $\Gamma$, the dark gray the best-fit
 values (Sec.~\ref{sec:Cacus}). The horizontal gray area is the Yarkovsky
 $\dd a / \dd t$ secular value from the orbit determination ($1\sigma$
 interval).}
\label{cacus_dadt}
\end{figure}

We also used optical astrometry of (161989)~Cacus
available to date and determined $\dd a / \dd t = -(1.6\pm 3.3)\times
10^{-4}$\,au\,Myr$^{-1}$. Unfortunately, there are no radar observations of Cacus and the optical data arc is not long enough yet to
reveal the Yarkovsky effect in Cacus' orbit. We note that our
value supersedes $\dd a / \dd t = (3.35\pm 2.3)\times 10^{-4}$\,au\,Myr$^{-1}$
from \cite{Far.ea:13} and $\dd a / \dd t = (2.6\pm 2.3)\times
10^{-4}$\,au\,Myr$^{-1}$ from \cite{Nug.ea:12}, both of which
are compatible with non-detection of the Yarkovsky effect. 
Figure~\ref{cacus_dadt} shows the computed $(\dd a / \dd t)_{\rm mod}$
values as a function of $\Gamma$. Therefore, the current
non-detection is very well explained by the small expected
$(\dd a / \dd t)_{\rm mod}$ value.

\subsection{(2100) Ra-Shalom}
In this case we used our derived rotation state parameters
from Sect.~\ref{sec:RaShalom} and assumed nominal size $D=2.3$\,km from
\cite{She.ea:08}. The spectral classification of this
object is somewhat unclear. \cite{Bin.ea:04} and \cite{Bus.Bin:02b}
classify it as a C- or Xc-type body,
but \cite{She.ea:08} give it K-type classification.
In this situation, we assume $1.7$\,g\,cm$^{-3}$ bulk density,
but if higher  or lower values turn out to be more appropriate
it is necessary to use the recalibration rules mentioned above.

With our nominal values we obtained $\upsilon_{\rm mod} \simeq
-10.0\times 10^{-8}\,\radd$. Interestingly, the much
reduced uncertainty interval for the pole orientation from
our solution in Sec.~\ref{sec:RaShalom} \citep[compare with][]{Dur.ea:12b}
now allows  us to consistently  predict
the deceleration of the rotation rate for this asteroid. This is
intriguing because it would be the first case of this sort.
We note, however, that  caution should be taken so as not to  jump to 
conclusions. First, the observations only allow us to constrain
$\upsilon$ to an interval that is compatible with a
zero value. Second, our simplified model seems to overpredict
the YORP strength by nearly an order of magnitude. This
may be partly due to the assumed low density or small size, but
also due to missing self-heating effects in our solution.
Additionally, the lateral conduction in surface irregularities
may be fine-tuned to  cancel the negative $\upsilon$
value from our model. We conclude that more observations are
needed to first set more meaningful constraints on $\upsilon$
in this complicated case before proceeding further with theoretical
implications.
\begin{figure}[pt]
\begin{center}
 \includegraphics[width=\columnwidth]{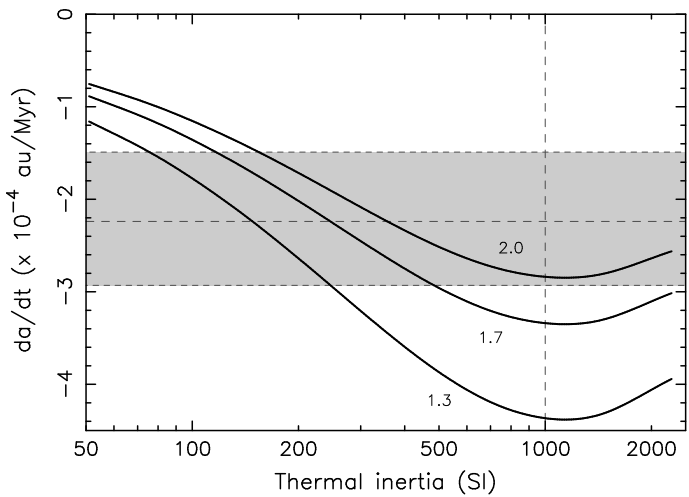}
\end{center}
\caption{ Same as in Fig.~\ref{toro_dadt}, but for asteroid
 (2100) Ra-Shalom. Here again, the Yarkovsky effect has been fairly
 well determined thanks to multitude of radar observations in
 five apparitions. The vertical dashed line indicates the estimated
 high thermal inertia of Ra-Shalom's surface from \cite{Del.ea:03} and 
 \cite{She.ea:08}. The three theoretical curves
 correspond to bulk densities of $1.3$, $1.7$, and $2$\,g\,cm$^{-3}$.}
\label{rasha_dadt}
\end{figure}

Unlike the YORP effect, the Yarkovsky effect has been fairly
well detected in the orbit of Ra-Shalom. This is again due to
very fortuitous circumstance of many radar observations, in this
case during five different close approaches to the Earth. Our
revision of the data yields $\dd a / \dd t = -(2.21\pm 0.72)\times
10^{-4}$\,au\,Myr$^{-1}$, somewhat smaller than the value  reported 
by \cite{Nug.ea:12} and \cite{Far.ea:13}. We
suspect the difference is due to new observations and 
the more recent statistical treatment for optical data we adopted here \citep{Far.ea:15, Ver.ea:17}.
Figure~\ref{rasha_dadt} shows that the agreement with
the predicted value $(\dd a / \dd t)_{\rm mod}$ is fairly good, provided
slightly higher than nominal density (and/or larger size) is assumed.
This may also help to alleviate the disagreement between the
estimated $\upsilon_{\rm mod}$ value and the limits on $\upsilon$
from observations.

\section{Conclusion}

In addition to the five asteroids for which YORP has already been detected: (1620)~Geographos, (1862)~Apollo, (3103)~Eger, (25143)~Itokawa, and (54509)~YORP \citep[see][and references therein]{Vok.ea:15}, we have another clear detection for (161989)~Cacus and a hint for YORP in asteroid (1685)~Toro. Another recent YORP detection is for asteroid Bennu (Nolan et al., 2017, ACM abstract). A striking feature of these detections is that all the $\dd \omega / \dd t$ values are positive, which means that for all these asteroids the rotation is accelerated. If there were the same number of asteroids with positive and negative $\upsilon$ values, the probability of all seven having the same sign just by chance would be $(1/2)^6 \simeq 1.6\%$ (or 0.8\% with eight of the same sign, if we include Toro). This low probability might mean that there is an asymmetry between accelerating and decelerating asteroid rotations with a preference for those that spin up. One of the possible mechanisms that would be consistent with this scenario is the transverse heat transport through surface boulders that always leads to acceleration \citep{Gol.Kru:12} and may have the same order of magnitude as the classical YORP \citep{Sev.ea:15}. However, this is still small-number statistics and our sample can be affected by selection bias. It is therefore crucial to significantly enlarge the sample of asteroids with a YORP detection.

A direct observational hint about the asymmetry
with which YORP prefers to accelerate the rotation of
small asteroids would have other interesting implications.
For instance, \cite{Pra.ea:08} \citep[see also updated data
in][]{Vok.ea:15} showed that small main-belt and
Hungaria asteroids have a rotation-rate distribution that is 
flat except for an overabundance of slow rotators. This
data set may be nicely explained with a YORP-relaxed population,
but the significant amount of slowly rotating bodies
requires that YORP also decelerates the rotation of asteroids. The
relative abundance of slow versus fast rotating bodies in the \cite{Pra.ea:08}
 model directly constrains how long asteroids remain
in the state of slow rotation before they re-emerge back
to regular rotation rates.

\begin{acknowledgements}
 This work was supported by the Czech Science Foundation (grants GA13-01308S, GA17-00774S, 
 and P209-12-0229). This publication also makes use of data products
  from NEOWISE, which is a project of the Jet Propulsion
  Laboratory/California Institute of Technology, funded by the
  Planetary Science Division of the National Aeronautics and Space
  Administration. D.~Farnocchia conducted this research at the Jet Propulsion Laboratory,
 California Institute of Technology, under a contract with NASA. The observations at Abastumani were supported by the Shota Rustaveli National Science Foundation, grant FR/379/6-300/14. The Cacus's lightcurve from 2003/02/18.1 was observed in cooperation with R.~Michelsen.
\end{acknowledgements}

\newcommand{\SortNoop}[1]{}

\end{document}